\documentclass[11pt,a4paper]{article}

\usepackage[hyperref]{acl2020}
\usepackage{times}
\usepackage{latexsym}

\usepackage{microtype}

\usepackage{amsmath}
\usepackage{mathtools}
\usepackage{graphicx}
\newcommand{\stcomp}[1]{\overline{#1}}
\DeclarePairedDelimiter{\ceil}{\lceil}{\rceil}

\def\hvec{\mathbf h}

\def\xvec{\mathbf x}

\def\yvec{\mathbf y}

\DeclareMathOperator*{\argmax}{arg\,max}

\newcommand\tokenfont[1]{{\usefont{T1}{custom}{m}{n}#1}}
\newcommand{\mask}[0]{$\langle \text{\tokenfont{MASK}} \rangle$}
\usepackage[ruled,vlined]{algorithm2e}

\aclfinalcopy 
\def\aclpaperid{2454} 

\title{Listen and Fill in the Missing Letters:\\Non-Autoregressive Transformer for Speech Recognition}

\author{Nanxin Chen \qquad Shinji Watanabe \qquad Jes\'us Villalba \qquad Najim Dehak \\
        Center for Language and Speech Processing \\
        Johns Hopkins University \\
        Baltimore, MD, USA
        }

\date{}

\begin{document}
\maketitle
\begin{abstract}
Recently very deep transformers have outperformed conventional bi-directional long short-term memory networks by a large margin in speech recognition.
However, to put it into production usage, inference computation cost is still a serious concern in real scenarios.
In this paper, we study two different non-autoregressive transformer structure for automatic speech recognition (ASR): A-CMLM and A-FMLM.
During training, for both frameworks, input tokens fed to the decoder are randomly replaced by special mask tokens.
The network is required to predict the tokens corresponding to those mask tokens by taking both unmasked context and input speech into consideration.
During inference, we start from all mask tokens and the network iteratively predicts missing tokens based on partial results.
We show that this framework can support different decoding strategies, including traditional left-to-right.
A new decoding strategy is proposed as an example, which starts from the easiest predictions to the most difficult ones.
Results on Mandarin (Aishell) and Japanese (CSJ) ASR benchmarks show the possibility to train such a non-autoregressive network for ASR.
Especially in Aishell, the proposed method outperformed the Kaldi ASR system and it matches the performance of the state-of-the-art autoregressive transformer with 7x speedup.
Pretrained models and code will be made available after publication.
\end{abstract}

\section{Introduction}
\label{sec:intro}

In recent studies, very deep end-to-end automatic speech recognition (ASR) starts to be comparable or superior to conventional ASR systems \citep{karita2019comparative,park2019specaugment,luscher2019rwth}.
They mainly use encoder-decoder structures based on long short-term memory recurrent neural networks \citep{chorowski2015attention,chan2016listen,watanabe2018espnet} and transformer networks \citep{dong2018speech,karita2019comparative,luscher2019rwth}.
Those systems have common characteristics: they rely on probabilistic chain-rule based factorization combined with left-to-right training and decoding.
During training, the ground truth history tokens are fed to the decoder to predict the next token.
During inference, the ground truth history tokens are replaced by previous predictions from the decoder.
While this combination allows tractable log-likelihood computation, maximum likelihood training, and beam-search based approximation, it is more difficult to perform parallel computation during decoding.
The left-to-right beam search algorithm needs to run decoder computation multiple times which usually depends on output sequence length and beam size.
Those models are well-known as auto-regressive models.

Recently, non-autoregressive end-to-end models have started to attract attention in neural machine translation (NMT)~\citep{gu2017non,lee2018deterministic,gu2019insertion,stern2019insertion,welleck2019non,ghazvininejad2019constant}.
The idea is that the system predicts the whole sequence within a \textit{constant} number of iterations which does not depend on output sequence length.
In \citep{gu2017non}, the author introduced hidden variables denoted as \emph{fertilities}, which are integers corresponding to the number of words in the target sentence that can be aligned to each word in the source sentence.
The fertilities predictor is trained to reproduce the predictions from another external aligner.
\citet{lee2018deterministic} used multiple iterations of refinement starting from some ``corrupted'' predictions.
Instead of predicting fertilities for each word in source sequence they only need to predict target sequence total length.
Another direction explored in previous studies is to allow the output sequence to grow dynamically \citep{gu2019insertion,stern2019insertion,welleck2019non}.
All those works insert words to output sequence iteratively based on certain order or explicit tree structure.
This allows arbitrary output sequence length avoiding deciding before decoding.
However since this insertion order or tree structure is not provided as ground truth for training, sampling or approximation is usually introduced to infer it.
Among all those studies of different directions, a common procedure for neural machine translation is to perform knowledge distillation \citep{gu2017non}.
In machine translation, for a given input sentence, multiple correct translations exist.
A pre-trained autoregressive model is used to provide a unique target sequence for training.

Our work is mainly inspired by the conditional language model proposed recently for neural machine translation \citep{ghazvininejad2019constant}.
Training procedure of this conditional language model is similar to BERT \citep{devlin2018bert}.
Some random tokens are replaced by a special mask token and the network is trained to predict original tokens.
The difference between our approach and BERT is that our system makes predictions conditioned on input speech.
Based on observations, we further propose to use factorization loss instead, which bridges the gap between training and inference.
During inference, the network decoder can condition on any subsets to predict the rest given input speech.
In reality, we start from an empty set (all mask tokens) and gradually complete the whole sequence.
The subset we chose can be quite flexible so it makes any decoding order possible.
In ASR, there is no need for knowledge distillation since in most cases unique transcript exists.

This paper is organized as follows.
Section~\ref{sec:at} introduces the autoregressive end-to-end model and section~\ref{sec:nat} discusses how to adapt it to non-autoregressive.
Different decoding strategies are also included.
Section~\ref{sec:exp} introduces the experimental setup and presents results on different corpora.
Further analysis is also included discussing the difference between autoregressive and non-autoregressive ASR.
Section~\ref{sec:con} summarizes this paper and provides several directions for future research in this area.

\section{Autoregressive Transformer-based ASR}
\label{sec:at}
 \begin{figure*}[tb]
     \centering
     \includegraphics[width=0.6\textwidth]{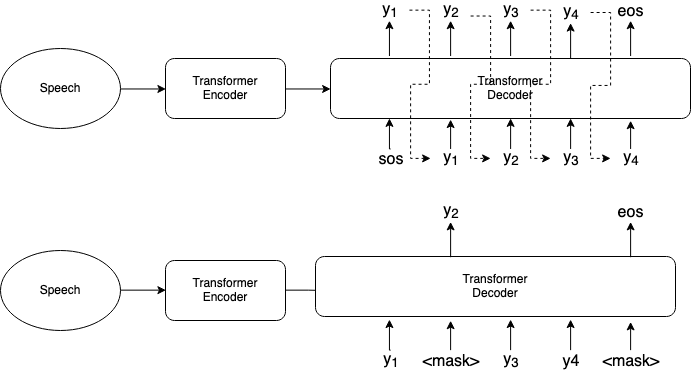}
     \caption{Comparison between normal transformer network and non-autoregressive transformer network. The transformer uses ground truth history tokens during training while during inference previous predictions are used as shown in the dash line. For non-autoregressive transformer training, random tokens in decoder input are replaced by a special \mask token and the network is required to predict for those positions. Both networks conditions on encoder outputs of the whole sequence.}
     \label{fig:comparison}
 \end{figure*}
To study non-autoregressive end-to-end ASR, it is important to understand how the current autoregressive speech recognition system works.
As shown in Figure~\ref{fig:comparison} top part, general sequence-to-sequence model consists of encoder and decoder.
The encoder takes speech features $\xvec_t$ like log Mel filter banks as input and produces hidden representations $\hvec _t$.
The decoder predicts a next token $\yvec _t$ based on the previous history $\yvec_{<t}$ and all hidden representations $\hvec = (\hvec_1, \hvec _2, \cdots)$:
\begin{equation}
    P(\yvec_t|\yvec_{<t}, \xvec) = P_{\mathsf{dec}}(\yvec_t | \yvec_{<t}, f_t(\hvec))
\label{eq:dec}
\end{equation}
where $f$ is a $t$-dependent function on all hidden representations $\hvec$.
A common choice for $f$ is an attention mechanism, which can be considered to be a weighted combination of all hidden representations:
\begin{equation}
    f^{\mathsf{att}}_t(\hvec) = \sum_{t^{'}} w_{t,t^{'}}\hvec_{t^{'}}
\end{equation}
where $t^{'}$ enumerates all possible hidden representations in $\hvec$. 
The weight $w_{t,t^{'}}$ is usually determined by a similarity between the decoder hidden state at $t$ and hidden representation $\hvec _{t'}$.

During training, the ground truth history tokens $\yvec_{<t}$ are usually used as input to the decoder in equation~\ref{eq:dec} for two reasons.
First, it is faster since the computation of all $P$ can be performed in parallel, as used in \citep{vaswani2017attention}.
Second, training can be very difficult and slow if predictions are used instead especially for very long sequence cases \citep{bengio2015scheduled,lamb2016professor}.
The expanded computation graph becomes very deep similar to recurrent neural networks without truncating.

During inference, since no ground truth is given, predictions need to be used instead.
This means equation~\eqref{eq:dec} needs to be computed sequentially for every token in output and each prediction needs to perform decoder computation once.
Depends on output sequence length and unit used, this procedure can be very slow for certain cases, like character-based Transformer models.

\section{Non-Autoregressive Transformer-based ASR}
\label{sec:nat}
Because of the training/inference discrepancy and sequential inference computation, non-autoregressive transformer becomes increasingly popular.

One possibility to make model non-autoregressive is to remove $\yvec_{<t}$ so parallel computation of $P(\yvec|\hvec)$ can be factorized as the product of $P(\yvec_t|\hvec)$.
However, this conditional independence might be too strong for ASR.
Instead, we replace $\yvec_{<t}$ with some other information like partial decoding results.
Similar to previous work \citep{lee2018deterministic, ghazvininejad2019constant}, multiple iterations are adopted to gradually complete prediction of the whole sentence.

\subsection{Training Frameworks}
\subsubsection{Audio-Conditional Masked Language Model (A-CMLM)}
One training framework we considered comes from \citet{ghazvininejad2019constant}.
The idea is to replace $\yvec_{<t}$ with partial decoding results we got from previous computations.
A new token \mask is introduced for training and decoding, similar to the idea of BERT \citep{devlin2018bert}.
Let $T_{\mathrm{M}}$ and $T_{\mathrm{U}}$ be the sets of masked  and unmasked tokens  respectively.
The posterior of the masked tokens given the unmasked tokens and the input speech is,
\begin{align}
    &P(\yvec_{T_\mathrm{M}}|\yvec_{T_\mathrm{U}},\xvec) = \prod_{t\in T_{\mathrm{M}}} P_{\mathsf{dec}}(\yvec_{t} | \yvec_{T_\mathrm{U}}, f_t(\hvec))\;.
\label{eq:mask}
\end{align}

As shown in Figure~\ref{fig:comparison} bottom part, during training some random tokens are replaced by this special token \mask.
The network is asked to predict original unmasked tokens based on input speech and context.
The total number of mask tokens is randomly sampled from a uniform distribution of whole utterance length and ground truth tokens are randomly selected to be replaced with this \mask token.
Theoretically, if we mask more tokens model will rely more on input speech and if we mask fewer tokens context will be utilized similar to the language model.
This combines the advantages of both speech recognition and language modeling.
We further assume that given unmasked tokens, predictions of masked tokens are conditionally independent so they can be estimated simultaneously as the product in equation ~\eqref{eq:mask}.

ASR uses audio as input (e.g., $f_t(
\hvec)$ in equation~\eqref{eq:mask}) instead of source text in NMT so we name this as audio-conditional masked language model (A-CMLM).

\subsection{Audio-Factorized Masked Language Model (A-FMLM)}
During the training of A-CMLM, ground truth tokens at $T_{\mathrm{U}}$ in~\eqref{eq:mask} are provided to predict the masked part.
However, during inference none of those tokens are given.
Thus the model needs to predict without any context information.
This mismatch can be arbitrarily large for some cases, like long utterances from our observations.
We will show it in the later experiments.

Inspired by \citet{yang2019xlnet, dong2019unified}, we formalize the idea to mitigate the training and inference mismatch as follows.
Let $Z_i \subset [0, 1, ..., T - 1]$ be a length-$(N + 1)$ sequence of indices such that
\begin{equation}
\begin{split}
    Z_0 & = \emptyset \\
    Z_N & = [0, 1, ..., T - 1] \\
     \forall i \quad Z_i & \subset Z_{i+1}
\end{split}
\end{equation}
For both training and inference the objective can be expressed as
\begin{equation}
    P(\yvec | \hvec) = \prod_{t=1}^N \prod_{i \in Z_t \cap {\stcomp Z_{t-1}}} P_{\mathsf{dec}}(\yvec_i | \yvec_{Z_{t-1}}, f_t(\hvec))
    \label{eq:a-fmlm}
\end{equation}
where $Z_t \cap {\stcomp Z_{t-1}}$ are the indices for decoding in iteration $t$.
For example, to decode utterance of length 5 with 5 iterations, one common approach (left to right) can be considered as:
\begin{equation}
    \begin{split}
        Z_0 & = \emptyset \\
        Z_1 & = {0} \\
        Z_2 & = {0, 1} \\
        Z_3 & = {0, 1, 2} \\
        Z_4 & = {0, 1, 2, 3} \\
        Z_5 & = {0, 1, 2, 3, 4} \\
    \end{split}
\end{equation}
Here $Z_t \cap {\stcomp Z_{t-1}} = t - 1$ so in this case equation \eqref{eq:a-fmlm} is equivalent to equation ~\eqref{eq:dec}.
Similar to A-CMLM, \mask tokens are added to decoder inputs when corresponding tokens are not decided yet.
The autoregressive model is a special case when $N=T$ and $Z_t=[0, 1, ..., t]$.

Ideally $Z_t$ should be decided based on confidence scores from the network predictions to match inference case.
During training we sort all posteriors from iteration $t-1$ and choose those most confident ones.
The size of $Z_t$ is also sampled from the uniform distribution between $0$ and $T$ to support different possibilities for decoding.
To speed-up, we set $N=2$ during training so that optimization objective can be also written as
\begin{equation}
\begin{split}
    P(\yvec | \hvec) = & \prod_{i \not\in Z_1} P_{\mathsf{dec}}(\yvec_i | \yvec_{t \in Z_1}, f_t(\hvec)) \\
    * & \prod_{j \in Z_1} P_{\mathsf{dec}}(\yvec_j | f_t(\hvec))
\end{split}
\end{equation}
Comparing with equation~\eqref{eq:mask}, A-CMLM training only includes first term if $Z_1 = T_{\mathrm{U}}$.
However, during inference, some explicit factorization is still needed.

Pseudo-code of our A-FMLM algorithm can be found in Algorithm~\ref{algo1}.

\begin{algorithm}[ht]
\SetAlgoLined
\SetKwInOut{Input}{input}
\SetKwInOut{Output}{output}
\Input{minibatch size $n$, dataset D, encoder network $f_{enc}$, decoder network $f_{dec}$}
\Output{Posterior $P$}
  Sample $x = \xvec_1$, $...$, $\xvec_n$, $y=\yvec_1$, $...$, $\yvec_n$ from D\;
  $\hvec = f_{enc}(\xvec)$\;
  Assign \mask to all elements in $\hat{\yvec}^0$\;
  $P(\yvec^1|\hvec)=f_{dec}(\hat{\yvec}^0, \hvec)$\;
  $mask = zeros(n, max\_length)$\;
  Assign \mask to all elements in $\hat{\yvec}^1$\;
  \For{i=$1$,$...$,$n$}{
      $probs = P_i(\yvec^1|\hvec)$\;
      $indices = argsort(probs.max(-1))$\;
      $Z \sim Uniform(1, length(probs))$\;
      $mask[i, indices[Z:]] = 1$\;
      $\hat{\yvec}^1[i, indices[Z:]] = y[i, indices[Z:]]$\;
  }
  $P(\yvec^2|\hvec)=f_{dec}(\hat{\yvec}^1, \hvec)$\;
  $P = mask * P(\yvec^1|\hvec) + (1 - mask) * P(\yvec^2|\hvec)$\;
 \caption{Minibatch forward pass}
 \label{algo1}
\end{algorithm}

It is also possible to introduce different networks $P_{\mathsf{dec}}$ for different iterations under our training framework.

\subsection{Decoding Strategies}
During inference, a multi-iteration process is considered.
Other than traditional left-to-right, two different strategies are studied: easy first and mask-predict.
\subsubsection{Easy first}
\label{sec:easy_first}
The idea of this strategy is to predict the most obvious ones first similar to easy-first parsing introduced in \citep{goldberg2010efficient}.
In the first iteration, the decoder is fed with predictions $\hat{\yvec}^0_t=\text{\mask}$  tokens for all $t$ since we do not have any partial results.
After getting decoding results $P(\yvec^1_t|.)$\footnote{we omit the dependecies of the posterior to keep the notation uncluttered}, we keep those most confident ones and update them in $\yvec_1$:
\begin{equation}
    \hat{\yvec}^1_t = \begin{cases}
\argmax_V P(\yvec^1_t|.) & t \in \mathrm{largest}_C(\max_V P(\yvec^1|.))\\
\hat{\yvec}^0_t &\text{otherwise}
\end{cases}
\end{equation}
where $V$ is the vocabulary, $C=\ceil{L/K}$ is the largest number of predictions we keep, $L$ is the sequence length and $K$ is the total number of iterations.
Conditioned on this new $\hat{\yvec}^1$, the network is required to make new predictions if there are still masked tokens.

Comparing with Algorithm~\ref{algo1}, we used predictions $\argmax_V P(\yvec^1_t|.)$ instead of ground truth tokens $y[i, indices[Z:]]$.

\subsubsection{Mask-predict}
This is studied in \citep{ghazvininejad2019constant}.
Similarly to Section~\ref{sec:easy_first}, we start with $\hat{\yvec}^0_t=\text{\mask}$.
In each iteration $k$, we check the posterior probability of the most probable token for each output $t$ (i.e., $\max_V{P(\yvec^k_t|.)}$) and use this probability as a confidence score to replace least confident ones in an utterance by \mask tokens.
The number of masked tokens in an utterance is $\ceil{L*(1-k/K)}$ for $k$-th iteration:
\begin{equation}
    {\footnotesize
    \hat{\yvec}^{k}_t = \begin{cases}
    \text{\mask} & t \in \mathrm{smallest}_C(\max_V{P(\yvec_t^k|.)})\\
    \argmax_V P(\yvec^{k}_t|.) & \text{otherwise}
    \end{cases}
    }
\end{equation}
where $C=\ceil{L*(1-k/K)}$.
For instance, if $K=10$, we mask $90\%$ tokens in the first iteration, $80\%$ in second and so on.
After getting prediction results we update all tokens previously masked in $\hat{\yvec}^{k-1}$:
\begin{equation}
    P(\yvec^{k}_t|.) = \begin{cases}
    P(\yvec^{k}_t|.) & \hat{\yvec}^{k-1}_t = \text{\mask}\\
    P(\yvec^{k-1}_{t}|.) & \text{otherwise}
    \end{cases}
\end{equation}
The difference between mask-predict and easy first is that mask-predict will accept all decisions but it reverts decisions made earlier if it is less confident.
Easy first is more conservative and it gradually adopts decisions with the highest confidence.
For both strategies, predictions become more and more accurate since it can utilize context information from both future and past directions.
This is achieved by replacing input $\yvec_{<t}$ with all $\yvec_t$ since left-to-right decoding is no longer necessary.

\subsection{Example}

  \begin{figure*}[tb]
     \centering
     \includegraphics[width=0.75\textwidth]{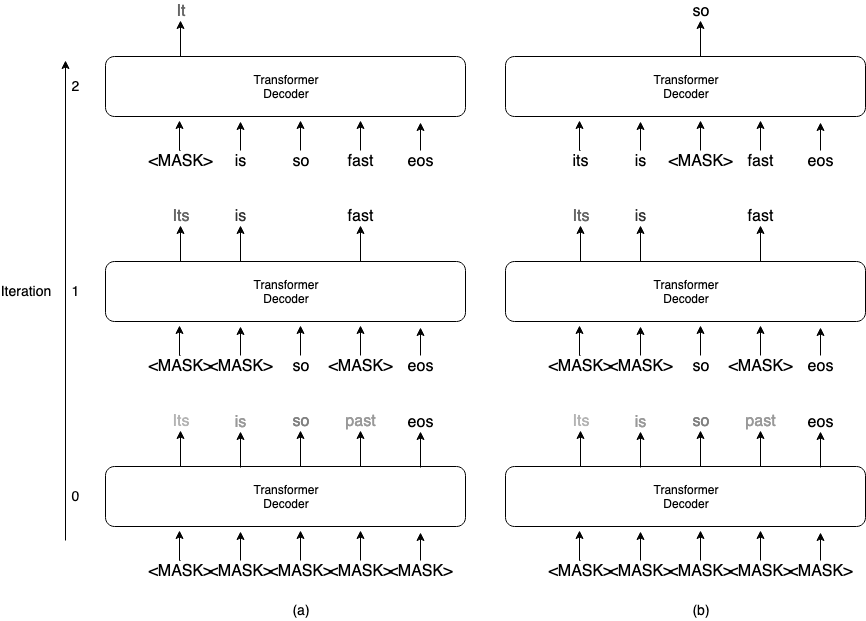}
     \caption{Illustration of inference procedure. To predict the whole sequence with $K=3$ passes, initially, the network is fed with all \mask tokens. Shade here presents the certainties from network outputs. Part (a) shows easy first process. Since token ``so'' is confident enough in the first iteration to be decided it will never change in the future. Part (b) shows mask-predict process. In the last iteration, it goes back to the word "so" because it is less confident in the first iteration compared to other predictions in other iterations. }
     \label{fig:mask_predict}
 \end{figure*}
 
One example is given in Figure~\ref{fig:mask_predict}.
Part (a) shows easy first and part (b) demonstrates mask-predict.
In this example sequence length is 4 but after adding $\langle \text{\tokenfont{EOS}} \rangle$ token to the end of the sequence we have $L=5$ and $K=3$.
In the first iteration, the network is inputted with all \mask.
Top $\ceil{5/3}=2$ tokens get kept in each iteration and based on partial results network predicts again on all rest \mask tokens.

For easy first, it always ranks confidence from the last iteration and then keep top-2 confident predictions.
Based on partial results it will complete the rest.

For mask-predict it maintains confidence scores from multiple iterations.
It chooses the least confident ones from all scores to mask.
In the last iteration, it chooses to change its previous prediction of ``so'' because its confidence is less than other predictions from the second iteration.

Normal inference procedure can be considered as a special case when $K=L$ and instead of taking the most confident one, the prediction of the next token is always adopted.
In general, this approach is flexible enough to support different decoding strategies: left-to-right, right-to-left, easy first, mask-predict and other unexplored strategies.

\subsection{Output sequence length prediction}
In \citet{ghazvininejad2019constant} they introduced a special token $\langle \text{\tokenfont{LENGTH}} \rangle$ in input to predict output sequence length.
For word sequence this is reasonable but for end-to-end speech recognition, it can be pretty difficult since character or BPE sequence length varies a lot.
In this paper a simpler approach is proposed: we asked the network to predict end-of-sequence token $\langle \text{\tokenfont{EOS}} \rangle$ at the end of the sequence as shown in Figure~\ref{fig:mask_predict}.

During inference, we still need to specify the initial length. We manually specify it to some constant value for the first iteration.
After that, we change it to the predicted length in the first iteration for speedup.


\section{Experiments}
\label{sec:exp}
For experiments, we mainly use Aishell \citep{bu2017aishell} and Corpus of Spontaneous Japanese(CSJ) \citep{maekawa2003corpus}.
Our preliminary investigations show that a Latin alphabet based task (e.g., English) has some difficulties since the mask-based character prediction in a word is easily filled out only with the character context without using any speech input (i.e., the network only learns the decoder).
Although the use of BPE-like sub-words can solve this issue to some extent, this introduces extra complexity and also the non-unique BPE sequence decomposition seems to cause some inconsistency for masked language modeling especially conditioned with the audio input. 
Due to these reasons, we chose ideogram languages (i.e., Mandarin and Japanese) in our experiments so that the output sequence length is limited and the prediction of the character is enough challenging. 
Applying the proposed method to the Latin alphabet language is one of our important future work.

ESPnet \citep{watanabe2018espnet} is used for all experiments. 
For the non-autoregressive baseline, we use state-of-the-art transformer end-to-end systems \citet{karita2019comparative}.
In Aishell experiments encoder includes 12 transformer blocks with convolutional layers at the beginning for downsampling.
The decoder consists of 6 transformer blocks.
For all transformer blocks, 4 heads are used for attention.
The network is trained for 50 epochs and warmup \citep{vaswani2017attention} is used for early iterations.
For all experiments beam search and language model is used and all configuration follows autoregressive baseline.

The results of Aishell is given in Table~\ref{tb:aishell}.
For A-CMLM, no improvement observed for more than 3 iterations.
For A-FMLM, experiments show that 1 iteration is enough to get the best performance.
Because of the connection between easy first and A-FMLM, we only use easy first for A-FMLM.
All transformer-based experiments are based on pure Python implementation so we don't compare them with C++-based Kaldi systems in the table.
It is still possible to get further speedup by improving current implementation.

All decoding methods result in performance very close to state-of-the-art autoregressive models.
Especially A-FMLM matched the performance of autoregressive baseline but real-time factor reduced from 1.44 to 0.22, which is around {\bf 7x speedup}.
The reason is that our non-autoregressive systems only perform decoder computation constant number of times, comparing to the autoregressive model which depends on the length of output sequence.
It also outperformed two different hybrid systems in Kaldi by 22\% and 11\% relative respectively.
\begin{table}[tb]
\centering
\resizebox{\columnwidth}{!}{
\begin{tabular}{cccc}\hline
{\bf System}                & \begin{tabular}[c]{@{}c@{}}{\bf Dev}\\{\bf CER}\end{tabular} & \begin{tabular}[c]{@{}c@{}}{\bf Test}\\{\bf CER}\end{tabular} & \begin{tabular}[c]{@{}c@{}}{\bf Real Time}\\{\bf Factor}\end{tabular} \\ \hline
Baseline(Transformer) & {\bf 6.0}                                         & {\bf 6.7}                                          &    1.44 \\
Baseline(Kaldi nnet3) & -                                                 & 8.6                                                &    - \\
Baseline(Kaldi chain) & -                                                 & 7.5                                                &    - \\ \hline
\citet{an2019cat}&-&6.3&-\\
\citet{fan2019unsupervised}&-&6.7&-\\\hline
Easy first(K=1)       & 6.8                                               & 7.6                                                &    0.22\\
Easy first(K=3)       & 6.4                                               & 7.1                                                &    0.22\\
Mask-predict(K=1)     & 6.8                                               & 7.6                                                &    0.22\\
Mask-predict(K=3)     & 6.4                                               & 7.2                                                &    0.24\\
A-FMLM(K=1)           & {\bf 6.2}                                         & {\bf 6.7}                                          &    0.28\\
A-FMLM(K=2)           & {\bf 6.2}                                         & 6.8                                          &    0.22\\ \hline
\end{tabular}}
\caption{Comparison of baselines, previous work, A-CMLM and A-FMLM on Aishell. For A-CMLM, easy first and mask-predict are compared. For A-FMLM, easy first is utilized since it connects to the factorization used in training.}
\label{tb:aishell}
\end{table}


CSJ results are given in Table~\ref{tb:csj}.
Here we observed a larger difference between non-autoregressive models and autoregressive models.
Multiple iterations of different decoding strategies are not helping to improve.
Still, A-FMLM we proposed outperforms A-CMLM with up to 9x speedup comparing to the autoregressive baseline.
\begin{table}[tb]
\centering
\resizebox{\columnwidth}{!}{
\begin{tabular}{ccccc}\hline
{\bf System}                & \begin{tabular}[c]{@{}c@{}}{\bf Eval1}\\ {\bf CER}\end{tabular} & \begin{tabular}[c]{@{}c@{}}{\bf Eval2}\\{\bf CER}\end{tabular}& \begin{tabular}[c]{@{}c@{}}{\bf Eval3}\\ {\bf CER}\end{tabular} &  \begin{tabular}[c]{@{}c@{}}{\bf Real Time}\\{\bf Factor}\end{tabular}  \\ \hline
Baseline(Transformer) & \textbf{5.9}                                              & \textbf{4.1}                                                &    \textbf{4.6} &  9.50\\
Baseline(Kaldi) & 7.5                                                 & 6.3                                                & 6.9   & - \\  \hline
Easy first(K=1)      &  8.8                                         & 6.7                                          & 7.4 & 1.31\\
Easy first(K=3)       & 9.3                                         & 7.0                                          &    8.3 & 1.05\\
Mask-predict(K=1)    &  8.8                                         & 6.7                                          & 7.4 & 1.31\\
Mask-predict(K=3)     & 11.7                                        & 8.8                                          &    9.9 & 1.01\\
A-FMLM(K=1)               & \textbf{7.7}                                          & \textbf{5.4}                                          &\textbf{6.2}    &  1.40 \\
A-FMLM(K=2)               & \textbf{7.7}                                          & \textbf{5.4}                                          &\textbf{6.2}    &  1.11 \\ \hline
\end{tabular}}
\caption{Comparison of baselines, previous work, A-CMLM and A-FMLM on CSJ. Same systems are compared as above.}
\label{tb:csj}
\end{table}

\begin{figure*}[htb]
     \centering
     \includegraphics[width=\textwidth]{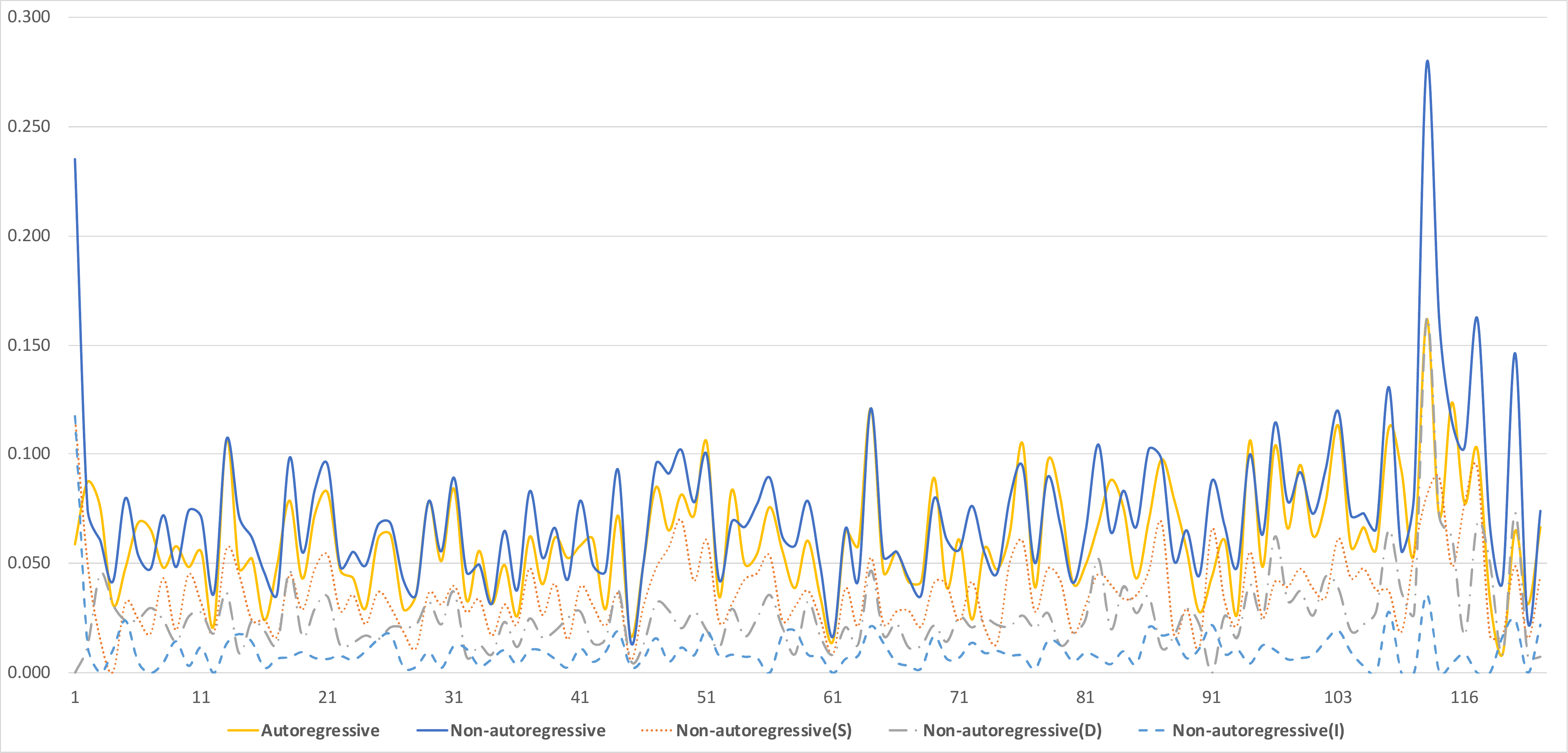}
     \caption{Error analysis of autoregressive and non-autoregressive on different output sequence length. Dash and dot lines indicate different errors: substitute(S), deletion(D), insertion(I)}
     \label{fig:cer}
\end{figure*}
To understand the performance difference between the autoregressive model and the non-autoregressive model, further analysis is included.
In Figure~\ref{fig:cer} we show the correlation between output sequence length and character error rate for Eval1 set.
For short utterances performance of the non-autoregressive model (blue line) is close to the autoregressive model (yellow line).
However, when output sequence becomes longer (large than 100), deletion error (long dash-dot line) and substitution error (dot line) start to increase dramatically.
This suggests the difficulty of handling very long utterances when the network is trained to predict all tokens simultaneously.
More specifically, error rate in the first iteration may increase since it is easy to miss certain token in the long sequence.
And those deletion errors increased discrepancy between training and inference which can not be easily fixed because all following tokens need to be moved.
This suggests the potential to apply insertion based models like \citet{gu2019insertion,stern2019insertion,welleck2019non}.

\section{Conclusion}
In this paper, we study two novel non-autoregressive training framework for transformer-based automatic speech recognition (ASR).
A-CMLM applies conditional language model proposed in \citep{ghazvininejad2019constant} to speech recognition.
Besides decoding strategies like left-to-right, mask-predict, a new decoding strategy is proposed.
Based on the connection with a classical dependency parsing \citep{goldberg2010efficient}, we named this decoding strategy easy first.
Inspired from easy first, we further propose a new training framework: A-FMLM, which utilizes factorization to bridge the gap between training and inference.
In experiments, we show that compared to classical left-to-right order these two show great speedup with reasonable performance.
Especially on Aishell, the speedup is up to 7 times while performance matches the autoregressive model.
We further analyze the problem of the non-autoregressive model for ASR on long output sequences.
This suggests several possibilities for future research, e.g., the potential to apply insertion based non-autoregressive transformer model to speech recognition.

\label{sec:con}



\bibliography{anthology,acl2020}
\bibliographystyle{acl_natbib}

\end{document}